\documentclass[slac_one]{revtex4}
\usepackage{graphicx}
\usepackage{fancyhdr}
\usepackage{epsfig,psfrag}

\begin{document}
\title{NNLO corrections to  $\bar B \to X_u  \ell \bar \nu $ 
and the determination of $|V_{ub}|$}

\author{Ben D. Pecjak}
\affiliation{Institut f\"ur Physik (THEP), Johannes Gutenberg-Universit\"at,
  D-55099 Mainz, Germany}

\begin{abstract}
The calculation of partial decay rates in $\bar B \to X_u  \ell \bar \nu $ 
decays at next-to-next-to-leading order (NNLO) in $\alpha_s$ and to leading
order in $1/m_b$ is described.  New results for the hard function are
combined with known results for the jet function and shape-function
moments in a numerical analysis which explores the impact of the NNLO
corrections on partial decay rates and the determination of $|V_{ub}|$.
\end{abstract}

\maketitle

\thispagestyle{fancy}

\section{INTRODUCTION} 
The inclusive decay $\bar B \to X_u  \ell \bar \nu $ is of much
interest because of its potential to constrain the CKM
element $|V_{ub}|$. Due to experimental cuts required to 
suppress charm background,  measurements of this decay are 
available only in the shape-function region, where the hadronic
final state is collimated into a single jet carrying a large 
energy on the order of $m_b$ and a moderate invariant 
mass squared on the order of $m_b \Lambda_{\rm QCD}$.  The theory 
challenge is to calculate partial decay rates in the presence of 
these cuts, where a local operator product expansion is insufficient.
A systematic treatment relies on a non-local operator product 
expansion whose end result can be formulated in terms of a 
factorization theorem.  Different approaches to this factorization
have been put forth in the 
literature, going under the names of BLNP \cite{Bosch:2004th,Lange:2005yw}, 
GGOU \cite{Gambino:2007rp}, and the dressed-gluon
exponentiation \cite{Andersen:2005mj} .  The purpose of this talk is 
to describe the elements that go into the BLNP formalism at 
next-to-next-to-leading order (NNLO) in 
$\alpha_s$ and leading order in $1/m_b$.

The main point of discussion is the factorization formula for an 
arbitrary decay distribution restricted to the kinematics of 
the shape-function region
\begin{equation}
\label{eq:VagueFact}
d\Gamma \sim H\cdot J\otimes S +{\cal O}\left(\frac{\Lambda_{\rm QCD}}{m_b}\right)\, ,
\end{equation}
where the symbol $\otimes$ denotes a convolution.
The factorization formula contains a hard function $H$, 
which is related to physics at the 
hard scale $m_b$, a  jet function $J$, which is related to physics at the
intermediate scale $m_b \Lambda_{\rm QCD}$, and a   
non-perturbative shape function $S$, describing
the internal soft dynamics of the $B$ meson 
\cite{Neubert:1993ch,  Bigi:1993ex}. This factorized form was originally
derived in \cite{Korchemsky:1994jb, Akhoury:1995fp} using diagrammatic
techniques, and was rederived in the framework of soft-collinear 
effective theory (SCET) in \cite{Bauer:2003pi, Bosch:2004th}.
The hard and jet functions to NLO in perturbation
theory have been known for some time \cite{Bauer:2003pi, Bosch:2004th}, 
and the NNLO jet function was obtained in \cite{Becher:2006qw}.  Very
recently, the NNLO contributions to the hard function have 
also been calculated \cite{Bonciani:2008wf,Asatrian:2008uk, Beneke:2008ei}, 
thus completing the perturbative corrections to the factorization
formula to this order.   In Section \ref{sec:NNLOfact} we review 
the calculation of $H$ and $J$ to NNLO, along with the form and 
solutions of their renormalization-group (RG) evolution equations.  
Then, in Section \ref{sec:Numerics}, we give preliminary
results illustrating the  numerical impact of the NNLO corrections 
on the $P_+$ spectrum,  comment on the relevance for the 
determination of $|V_{ub}|$, and make some concluding remarks.

\section{THE HARD AND JET FUNCTIONS AT  NNLO}
\label{sec:NNLOfact}
In this section we describe the calculation of the hard and jet
functions at NNLO, as well as the form and solution
of their RG evolution equations.

The hard function arises when integrating out fluctuations at the 
scale $m_b$ by matching QCD onto SCET.  This matching can be done
at the level of the $b\to u$ transition current, and to leading
order in $1/m_b$ takes the form  
\begin{equation}
\label{eq:PosMatching}
e^{-im_b v \cdot x} \bar u(x)\gamma^\mu (1-\gamma_5)b(x) =
\sum_{i=1}^3 \int ds \, \tilde C_i(s)\bar \chi(x+s\bar n)\Gamma^\mu_i {\cal H}(x_-),
\end{equation} 
where we have followed the SCET conventions of \cite{Bosch:2004th}, and  
the $\Gamma^\mu_i$ are a set of three Dirac structures.  
In practice, the matching calculation is carried out in momentum space and 
yields results for the Fourier-transformed coefficients, which read
\begin{equation}
C_i(\bar n \cdot p) = \int  ds \, e^{is\bar n\cdot p} \, \tilde C_i(s)\,.
\end{equation}
The hard function $H$ is derived from the matrix of coefficients
$H_{ij}=C_i C_j$.

To obtain the matching coefficients $C_i$ requires to calculate
UV-renormalized matrix elements in full QCD and SCET.  The calculation 
is simplest when the external states are chosen as on-shell quarks
and both UV and IR divergences are regulated in dimensional regularization
in $d=4-2\epsilon$ dimensions.  In that case the loop corrections
to the SCET matrix elements are given by scaleless integrals and vanish,
so that the result is just its tree-level value
multiplied by renormalization factors from operator and 
wave-function renormalization. The main challenge is to calculate
the QCD result, which is written
in terms of three Dirac structures multiplied by scalar form factors 
$D_{i}$.  To obtain these scalar form factors at NNLO requires 
to calculate the two-loop corrections to the $b\to u$ current in QCD.
This task has been completed in 
\cite{Bonciani:2008wf,Asatrian:2008uk, Beneke:2008ei}.   
One then turns the results for the scalar 
form factors $D_i$ into those 
for the SCET Wilson coefficients by evaluating the matching conditions    
\begin{eqnarray}
\label{eq:Matching}
C_i(m_b,\bar n \cdot p,\mu) &=&   \lim_{\epsilon\to 0}\,Z_J^{-1}(\epsilon,m_b, \bar n \cdot p, \mu)
D_i(\epsilon,m_b, \bar n \cdot p, \mu) \, ,
\end{eqnarray}
where $Z_J$ is a current renormalization factor.  Details of the 
matching calculation and results for the Wilson coefficients 
have been presented in \cite{Asatrian:2008uk, Beneke:2008ei}.

In the BLNP approach, the hard function is evaluated at a scale 
$\mu_i\sim 1.5$~GeV and logarithms  of the ratio $\mu_h/\mu_i$ 
are treated as large. One can resum these logarithms by deriving
and solving the RG equation for the hard coefficients $C_i$, or 
equivalently the matrix of coefficients $H_{ij}$.  The evolution
equation has the form (see, e.g. \cite{Bosch:2004th})
\begin{equation}
\label{eq:RGH}
\frac{d}{d\ln \mu} H_{ij}
(\bar n \cdot p, \mu)=
2\left[
\gamma^\prime(\alpha_s)+\Gamma_{\rm cusp}(\alpha_s)\ln\frac{\bar n \cdot p}{\mu}\right]H_{ij}(\bar n \cdot p,\mu) \,,
\end{equation} 
and its solution is 
\begin{equation}
\label{eq:resummedh}
H_{ij}(\bar n \cdot p,\mu)=y^{-2a_\Gamma(\mu_h,\mu)} {\rm exp}\left[2 S(\mu_h,\mu)-2 a_\Gamma(\mu_h,\mu)\ln\frac{m_b}{\mu_h}-2a_{\gamma^\prime}(\mu_h,\mu)\right]
H_{ij}(\bar n \cdot p,\mu_h)\,,
\end{equation}
where we have defined $y=\bar n \cdot p/m_b$.
Explicit results for the Sudakov factor $S$ and the anomalous 
exponents $a_\Gamma$, $a_{\gamma^\prime}$ can be read off 
from \cite{Bosch:2004th}.  A consistent treatment at 
NNLO in $\alpha_s$ requires 
the matching coefficients at two loops, the anomalous dimension 
$\gamma^\prime$ at three loops, and the cusp anomalous dimension at four 
loops. The anomalous dimensions are both known to one loop lower, which 
adds a small uncertainty to the analysis.  

We now turn to the jet function $J$, which
arises when integrating out fluctuations at the 
intermediate scale $m_b \Lambda_{\rm QCD}$ by matching SCET onto 
HQET.  The calculation of the jet function can be recast into the 
evaluation of the imaginary part of a certain vacuum matrix 
element in QCD, and was calculated to NNLO in  \cite{Becher:2006qw}.
In light-cone gauge, this matrix element is the quark propagator.  
The jet function depends on a renormalization scale $\mu$, which is 
usually kept fixed at $\mu_i=1.5$~GeV in the BLNP approach.
However, it is possible to derive and solve the RG equation for $J$, which
allows one to also vary the renormalization scale at which the jet 
function is evaluated.   The starting point is the exact 
integro-differential evolution 
equation derived in  \cite{Becher:2006qw}:
\begin{equation}\label{Jrge}
   \frac{dJ(p^2,\mu)}{d\ln\mu}
   = - \left[ 2\Gamma_{\rm cusp}\,\ln\frac{p^2}{\mu^2}
+ 2\gamma^J \right] J(p^2,\mu)
   - 2\Gamma_{\rm cusp} \int_0^{p^2}\!dp^{\prime 2}\,
   \frac{J(p^{\prime 2},\mu)-J(p^2,\mu)}{p^2-p^{\prime 2}} \,.
\end{equation} 
A novel solution to the evolution equation (\ref{Jrge}) 
was presented in \cite{Becher:2006nr}.  
The method is based on the observation that the 
Laplace transformed jet function 
\begin{equation}\label{Laplace}
   \widetilde j\Big( \ln\frac{Q^2}{\mu^2},\mu \Big)
   = \int_0^\infty\!dp^2\,e^{-s p^2}\,J(p^2,\mu) \,, \qquad
   s = \frac{1}{e^{\gamma_E} Q^2} \,
\end{equation}
obeys a local evolution equation analogous to 
(\ref{eq:RGH}).  Solving as before and inverting the 
Laplace transform gives the resummed
$J$ in momentum space. Alternatively, it is always possible
to write partial decay rates in terms of an integral over the jet 
function, defined as 
\begin{equation}
\label{eq:jdef}
j\left(\ln\frac{Q^2}{\mu^2},\mu\right)=\int_0^{Q^2} dp^2 \, J(p^2,\mu) \, .
\end{equation}
Using the solution of the RG equation, one can write a compact expression
for the resummed integrated jet function
\begin{equation}
\label{eq:resummedj}
j\Big(\ln\frac{Q^2}{\mu_f^2},\mu_f\Big)
={\rm exp}\left[-4 S(\mu_i,\mu_f)+2 a_{\gamma^J}(\mu_i,\mu_f)\right]
\left(\frac{Q^2}{\mu_i^2}\right)^\eta
\widetilde j\Big(\partial_\eta+\ln\frac{Q^2}{\mu_i^2},\mu_i\Big)
\frac{e^{-\gamma_E\eta}}{\Gamma(1+\eta)} ,
\end{equation}
where expressions for the RG factors $a_{\gamma^J}$ and $\eta$ can 
be found in \cite{Becher:2006nr}.
Equations (\ref{eq:resummedh})  and (\ref{eq:resummedj}) allow one 
to evaluate integrals of the product $H\cdot J$ at an arbitrary 
scale $\mu_f$ at which the shape function is renormalized, 
and study the dependence
of partial decay rates under independent variations of the matching
scales $\mu_h$ and $\mu_i$, under which they are formally 
independent.  Detailed numerical results shall be given in \cite{asatrian}.

\section{NUMERICAL RESULTS AND CONCLUSIONS}
\label{sec:Numerics}
%%%%%%%%%%%%%%%%%%%%%%%%%%%%%%%%%%%%%%%%%%%%%%%%%%%%
\begin{figure*}[t]
\centering
\psfrag{x}[]{$\mu_h$ (GeV)}
\psfrag{left}[]{ $\frac{\Gamma_u(P_+<\Delta)}{|V_{ub}|^2 {\rm ps}^{-1}}$}
\psfrag{prelim}[]{}
\psfrag{lo}[]{{\footnotesize LO}}
\psfrag{nlo}[]{{\footnotesize NLO}}
\psfrag{nnlo}[]{{\footnotesize NNLO}}
\includegraphics[width=0.4 \textwidth]{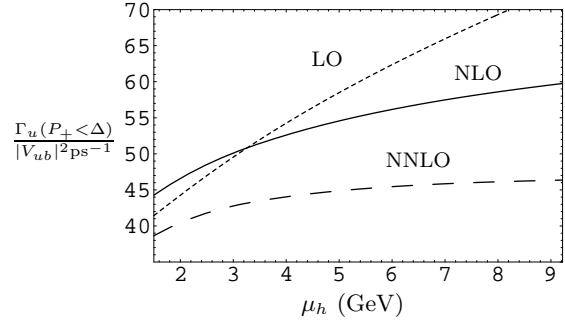}
\caption{Partial rates with a cut $P_+<M_D^2/M_B$ as a function 
of $\mu_h$ at LO, NLO, and NNLO.} \label{fig:ScaleVar}
\end{figure*}
%%%%%%%%%%%%%%%%%%%%%%%%%%%%%%%%%%%%%%%%%%%%%%%%%%%%
In this section we briefly explore the numerical impact of the NNLO
corrections, taking as an example the partial rate with a 
cut $P_+<\Delta=M_D^2/M_B$.  This partial rate can 
be written as 
\begin{eqnarray}
\label{eq:P+} 
&&   \Gamma_u(P_+<\Delta)
   = \frac{G_F^2|V_{ub}|^2}{96\pi^3}
\int_0^\Delta d\hat\omega \, \hat S(\hat\omega,\mu_f) 
\int_0^{M_B-\hat\omega} \,dq \,5q^4
%\\ \nonumber 
%&& 
%\hspace{0.2cm}
\int_0^{1}\!dy\,y^{2-2a_\Gamma} H_u(y,\mu_f)
j\left(\ln\frac{m_by(\Delta_q-\hat\omega)}{\mu_f},\mu_f \right)  
\end{eqnarray}
where $\Delta_q={\rm min}(\Delta, M_B-q)$, and 
the hard and integrated jet functions are to be evaluated 
in their resummed forms  (\ref{eq:resummedh})  and (\ref{eq:resummedj}).
 The function $H_u$ 
is related to certain combinations of the $H_{ij}$ and can be 
deduced from \cite{Lange:2005yw}.
To evaluate (\ref{eq:P+}) requires a model for the shape function.
In what follows, we use the two-parameter exponential model
\begin{equation}
\label{eq:Smod}
\hat S(\hat \omega)={\cal N}(b,\Lambda)\hat \omega^{b-1} 
{\rm exp}\left(-\frac{b\hat \omega}{\Lambda}\right) \, .
\end{equation}
It is possible to put model-independent constraints on the 
normalization factor ${\cal N}$ and the parameters 
$b$ and $\Lambda$ by studying shape-function moments \cite{Bosch:2004th}.  
 These moments are defined as
\begin{equation}\label{Def:Moments}
   M_N(\hat\omega_0,\mu_i)\equiv \int_0^{\hat\omega_0}\!d\hat\omega\,
   \hat\omega^N\,\hat S(\hat\omega,\mu_i) \,,
\end{equation}
and can be obtained as an expansion in local HQET operators
as long as $\hat\omega_0\gg \Lambda_{\rm QCD}$. The zeroth moment
fixes the normalization of the shape function and was 
determined to NNLO in  \cite{Becher:2005pd}, and can be used along
with the first and second moment to define the parameters 
$m_b = M_B - \bar \Lambda$ and $\mu_\pi^2$ in the shape-function scheme. 
Numerical values for 
the HQET parameters $m_b$ and $\mu_\pi^2$ in the shape-function 
scheme at NNLO can be determined from those obtained from global
fits in other schemes using the two-loop conversion relations
derived in  \cite{Neubert:2004sp}.  One then tunes the parameters
$b$ and $\Lambda$ such that the moments of the model function
(\ref{eq:Smod}) give appropriate results for $m_b$
and $\mu_\pi^2$.  In our analysis here, we tune the parameters such that  
$m_b=4.61$~GeV and $\mu_\pi^{2}=0.2\, {\rm GeV}^2$ 
at NNLO in the shape-function scheme.

In Figure \ref{fig:ScaleVar}, we show results for the partial rate 
with a cut $P_+<\Delta=M_D^2/M_B$, for the choice $\mu_i=\mu_f=1.5$~GeV,
as a function of the hard matching scale $\mu_h$.  
From the figure, one sees that the dependence on the renormalization scale
$\mu_h$ is reduced when going to higher orders in perturbation theory,
and also that there is a fairly large downward shift in the central 
value between NLO and NNLO.  This would tend to raise the value of $|V_{ub}|$
compared to the result deduced from the NLO calculation in the BLNP approach.  
However, we must  stress that these numerical results are preliminary, 
and will be finalized in \cite{asatrian}, along with uncertainties 
associated with variations of the matching scale $\mu_i.$  Results for 
other partial rates, such as those with cuts on the hadronic invariant mass and
lepton energy, will also be presented.

It is worth emphasizing that the partial decay rates are rather 
sensitive to the numerical value of $m_b$.  If, as suggested in
\cite{Neubert:2008cp}, the 
$B\to X_s\gamma$ moments are excluded from global fits on the grounds
that measurements of the photon energy spectrum are made in kinematic regions
where shape-function effects are important, then $m_b$ tends to come out 
closer to $4.71$~GeV.  Changing $m_b$ by such an amount raises the partial
decay rates upwards by approximately 20\%. A precise determination of 
$|V_{ub}|$ from inclusive decays will thus require that this point be settled.

\begin{acknowledgments}
I am grateful to Hrachia Asatrian, Christoph Greub, and Matthias Neubert,
for collaborations on the work which formed the basis of this talk.
\end{acknowledgments}

%\begin{thebibliography}{9}   % Use for  1-9  references

\end{document}